\documentclass[pra,onecolumn]{revtex4}
\usepackage{epsfig}
\usepackage{amssymb}

\newtheorem{corollary}{Corollary}
\newtheorem{theorem}{Theorem}
\newtheorem{lemma}{Lemma}
\newtheorem{property}{Property}
\newcommand{\Tr}{\textrm{Tr}}

\begin{document}

\title{Optimal Unambiguous State Discrimination of two density matrices \\
and its link with the Fidelity}

\author{Philippe Raynal}

\author{Norbert L\"utkenhaus}

\affiliation{Quantum Information Theory Group, Institut f\"ur Theoretische Physik I, and \\
 Max-Plank-Forschungsgruppe, Institut f\"ur Optik, Information und Photonik\\ Universit\"at
Erlangen-N\"urnberg, Staudtstr. 7, D-91058 Erlangen, Germany}

\date{\today}
\begin{abstract}

Recently the problem of Unambiguous State Discrimination (USD) of mixed quantum states has attracted much attention. So far, bounds on the optimum success probability have been derived \cite{rudolph03a}. For two mixed states they are given in terms of the fidelity.  Here we give tighter bounds as well as necessary and sufficient conditions for two mixed states to reach these bounds. Moreover we construct the corresponding optimal measurement strategies. With this result, we provide analytical solutions for unambiguous discrimination of a class of generic mixed states. This goes beyond known results which are all reducible to some pure state case. Additionally, we show that examples exist where the bounds cannot be reached. 
\end{abstract}

\maketitle

\section{Introduction}

Quantum state discrimination \cite{helstrom76a} is a fundamental task in quantum information theory, especially in a communication context. Whenever the signal states are nonorthogonal, perfect discrimination becomes impossible. One has then to resort to optimum state discrimination strategies by specifying figures of merit that define some optimal strategies. The optimum strategy depends then on the quantum states and their {\it a priori} probabilities. One strategy is {\em Minimum Error Discrimination} (MED) \cite{helstrom76a} in which the measurement identifies the possible input states with some error. It is the goal to minimize the error. Another strategy is to optimize the mutual information between the sender and receiver. 

The scenario studied here is  {\em Unambiguous State Discrimination} (USD) which characterizes a measurement which either identifies a signal state without error ('unambiguous') or sends out a flag stating that it failed to identify the state. The objective is to minimize this failure probability.  The problem of finding optimal USD strategies has been solved for many pure state
scenarios \cite{dieks88a,ivanovic87a,peres88a,peres98a,chefles98a}, including any two pure states \cite{jaeger95a}.

In contrast to the  MED problem, which is already solved for any pair of mixed states \cite{helstrom76a}, optimal USD of mixed states is an open problem. Some special cases have been given for which the corresponding problem can be reduced to USD of pure state case, such as in  state filtering \cite{sun02a, bergou03a,herzog05a} or state comparison \cite{rudolph03a,herzog05a}. The underlying  reduction theorems have been stated in \cite{raynal03a}. For the general case,  necessary and sufficient conditions for the optimality of a POVM were derived in \cite{fiurasek03a, eldar04a}. They allow a numerical treatment of the problem but have not given rise to analytic solutions.  

For the unambiguous discrimination of a pair of mixed states, lower bounds on the failure probability have been found \cite{rudolph03a, feng04a} and reveal three regimes, depending on the ratio between the two {\it a priori} probabilities of the two mixed states. The boundaries of the middle regime were recently refined in \cite{herzog05a} but the consequences for the two remaining outer regimes were not addressed. Here we provide new bounds in those two regimes. Furthermore we derive necessary and sufficient conditions to reach the three bounds in the three different regimes. Given two density matrices $\rho_0$ and $\rho_1$ and their {\em a priori} probabilities $\eta_0$ and $\eta_1$, the necessary and sufficient conditions to reach the bounds given here take the form of the  positivity of two particular operators. Moreover we show that examples exist where the bounds cannot be attained. When the necessary and sufficient conditions are fulfilled, we give the optimal measurement strategy to reach the bounds.

The structure of this paper is the following. In the Sec. \ref{bounds}, we derive lower bounds for the success probability in the case of two mixed states. Our  derivation uses the Cauchy-Schwarz inequality, as used in \cite{barnum96a}, and  allows us to look for necessary and sufficient conditions to reach the lower bound in each regime of the {\em a priori} probabilities. In Sec. \ref{sec:parallel}, we report the notion of {\em parallel addition} that leads to some useful relations for USD in connection with a reduction theorem of Ref.\cite{raynal03a}. In Sec.\ref{conditions}, we derive the main result of this paper as a theorem: two necessary and sufficient conditions for the failure probability to reach the bounds are given. We also give the corresponding optimal POVM. In Sec.\ref{discussion}, we provide examples showing that there are generic mixed states of interest for which the necessary and sufficient conditions are fulfilled and for which we can therefor give the optimal USD measurement.

\section{Lower bounds on the failure probability}
\label{bounds}
	 
In Unambiguous State Discrimination, the performed measurement either identifies uniquely a state (conclusive result) or fails to identity it (inconclusive result). The goal is to optimize that strategy by finding the measurement for which the probability of inconclusive result is as small as possible. The problem is then specified by the set of quantum signal states $\{\rho_i\}$ and their respective {\it a priori} probabilities $\{\eta_i\}$. The measurement is a generalized measurement i.e.\ a Positive Operator-Valued Measure (POVM) \cite{helstrom76a}. A POVM is a set of hermitian  and positive semi-definite  operators $\{E_i\}_i$ that add up to identity acting on the Hilbert space spanned by the signal states, i.e.\ $\sum_i E_i=\openone_{\mathcal H}$. Given $N$ possible input states, we consider measurements with $N+1$ outcomes where the first $N$ outcomes identify a state and the last one corresponds to inconclusive results where the identification failed. The POVM elements are denoted by $E_k$ with $k=1,\dots,N$ and $E_?$ respectively.  The probability to obtain an outcome for some POVM element $E$ for a given signal $\rho$ is then given by $\Tr(E \rho)$. 

In general, a POVM describing a USD measurement satisfies $\Tr(E_k \rho_i)=0$ whenever $k\neq i$ so that only the state $\rho_i$ can trigger the measurement outcome connected to $E_k$. The failure probability $Q$ of a USD strategy is then given by $Q=\sum_i Q_i$, where $Q_i= \sum_i \eta_i \Tr (E_? \rho_i)$. From this definition we find that $Q_i \le \eta_i$. In this paper, we consider the USD of two signal states $\rho_0$ and $\rho_1$ that are mixed states with {\it a priori} probabilities $\eta_0$ and $\eta_1$. Accordingly, our POVM contains three elements $\{E_0,E_1,E_?\}$ which correspond respectively to the conclusive detection of $\rho_0$, to the conclusive detection of $\rho_1$ and to an inconclusive result. The failure probability then equals $Q=Q_0+Q_1$.

Our interest is first focused on the product $Q_0Q_1$. We can give a lower bound that is expressed in terms of the fidelity $F$ of the two states. The fidelity is defined as $F = \Tr(\sqrt{\sqrt{\rho_0}\rho_1\sqrt{\rho_0}})$ \cite{jozsa94a}. The bounds, formulated in the following theorem, are tighter than the one given in \cite{rudolph03a}. Moreover, we pay additionally attention to the condition under which the bound can be reached. 

As for the notation, consider an hermitian and positive semi-definite operators $O$. We can define its unique square root $\sqrt{O}$ and decompose it into the form $O=MM^{\dagger}$ with $M=\sqrt{O}U$, for any unitary matrix $U$. Since the states $\rho_i$ and the POVM elements $E_k$ all are hermitian and positive semi-definite operators, we can introduce their square root and  use the previous decomposition.

\begin{theorem}
Let $\rho_0$ and $\rho_1$ be two density matrices with {\it a priori} probabilities $\eta_0$ and $\eta_1$. We define the fidelity of the two states $\rho_0$ and $\rho_1$ as $F= \Tr(\sqrt{\sqrt{\rho_0}\rho_1\sqrt{\rho_0}})$. Then, for any USD measurement, the product of the two probabilities $Q_0$ and $Q_1$ to fail to identify respectively the state $\rho_0$ and $\rho_1$ is such that
\begin{eqnarray}
Q_0 Q_1\ge \eta_0 \eta_1F^2.
\end{eqnarray}
The equality holds if and only if the unitary operator $V$ arising from a polar decomposition
\begin{eqnarray}
\sqrt{\rho_0}\sqrt{\rho_1}=\sqrt{\sqrt{\rho_0}\rho_1\sqrt{\rho_0}}\,\, V \nonumber
\end{eqnarray}
satisfies
\begin{equation} \label{condition}
V^\dagger \sqrt{\rho_0}E_?=\alpha \sqrt{\rho_1} E_?\\
\end{equation}
for some $\alpha \in \mathbb{R}^+$.
\end{theorem}
Before we turn to the proof of this theorem note that relation (\ref{condition}) implies a condition required for the optimality of a USD POVM (see \cite{bergou03a, raynal03a}). It is clear that optimality of a specific USD measurement implies that the conditional states after the inconclusive results do not allow further USD measurements. That would already be satisfied if, for example, the supports of the conditional states coincide. We find a stronger property whenever equality holds in  Theorem 1. Indeed, if  we have $V^\dagger \sqrt{\rho_0}E_?= \alpha \sqrt{\rho_1} E_?$  with $\alpha \in \mathbb{R}^+$, then it follows immediately that  $\sqrt{E_?} \rho_0 \sqrt{E_?} = \alpha^2 \sqrt{E_?} \rho_1 \sqrt{E_?}$. This means that the conditional states corresponding to inconclusive results must be identical up to normalization. Therefor no information whatsoever about the signal state can be extracted from these conditional states.

\paragraph*{\bf Proof of Theorem 1}
The basic ingredient for the derivation of the bound is the Cauchy-Schwarz inequality:

\begin{theorem} \cite{lancaster85}
Cauchy-Schwarz inequality\\
If x and y are members of a unitary space then
$\|x\| \|y\|\ge |(x,y)|$.
The equality holds if and only if $x=\alpha \, y$ for some $\alpha$ in $\mathbb C$.
\end{theorem}
A unitary space is  a complex linear space $\mathcal{S}$ together with an inner product from $\mathcal{S} \times \mathcal{S}$ to $\mathbb C$. Therefore the complex space of bounded operators acting on a Hilbert space is a complete unitary space if we consider for two elements $A, B$ the inner product $\Tr(A B^{\dagger})$. Hence, with $E_i=M_iM_i^{\dagger}$, $\rho_0=\sqrt{\rho_0}U U^\dagger \sqrt{\rho_0}$ and $\rho_1=\sqrt{\rho_1}\sqrt{\rho_1}$, we obtain
\begin{eqnarray*}
\sqrt{\Tr(E_i \rho_0)} \sqrt{ \Tr(E_i \rho_1)} &=& \sqrt{\Tr(U \sqrt{\rho_0}M_i M_i^\dagger \sqrt{\rho_0}U^\dagger)} \sqrt{\Tr(\sqrt{\rho_1} M_i M_i^\dagger \sqrt{\rho_1})}  \\
&\ge& |{\Tr(U \sqrt{\rho_0} M_i M_i^\dagger \sqrt{\rho_1})|},
\end{eqnarray*}
where we have used the freedom in the decomposition of $\rho_0$. By Theorem 2, the equality holds if and only if $U \sqrt{\rho_0}M_i=\alpha \sqrt{\rho_1} M_i$, for some $\alpha \in \mathbb{C}$ or, equivalently, if and only if $U \sqrt{\rho_0}E_i=\alpha \sqrt{\rho_1} E_i$, for some $\alpha \in \mathbb{C}$.
\\
We now consider a USD POVM $\{E_i\}_{i=0,1,?}$. Using the fact that $\Tr(E_0 \rho_1)=\Tr(E_1 \rho_0)=0$, we find for $E_0$ and $E_1$

\begin{eqnarray*}
0=\sqrt{\Tr(E_0 \rho_0)} \sqrt{\Tr(E_0 \rho_1)} \ge |\Tr(U \sqrt{\rho_0} E_0 \sqrt{\rho_1})|,\\
0=\sqrt{\Tr(E_1 \rho_0)} \sqrt{\Tr(E_1 \rho_1)} \ge |\Tr(U \sqrt{\rho_0} E_1 \sqrt{\rho_1})|.
\end{eqnarray*}
This simply means that $\Tr(U \sqrt{\rho_0} E_0 \sqrt{\rho_1})=\Tr(U \sqrt{\rho_0} E_1 \sqrt{\rho_1})=0$.
For $E_?$, we obtain
\begin{eqnarray*}
\sqrt{\Tr(E_? \rho_0)} \sqrt{\Tr(E_? \rho_1)}\ge |\Tr(U \sqrt{\rho_0} E_? \sqrt{\rho_1})|.
\end{eqnarray*}
From this it follows that  we can write
\begin{eqnarray} \label{CS}
\sqrt{\Tr(E_? \rho_0)} \sqrt{\Tr(E_? \rho_1)}\ge |\Tr(U \sqrt{\rho_0} E_? \sqrt{\rho_1})+0+0|=|{\Tr(U \sqrt{\rho_0}\sqrt{\rho_1})|}\;,
\end{eqnarray}
where we used the relation $\sum_i E_i = \openone$. Furthermore, the inequality (\ref{CS}) must hold for any unitary matrix $U$ so that we find
\begin{equation} \label{max}
\sqrt{\Tr(E_? \rho_0)} \sqrt{\Tr(E_? \rho_1)}\ge \max_U |\Tr(U \sqrt{\rho_0} \sqrt{\rho_1})|.
\end{equation} Here, again, the equality holds if and only if a unitary operator $U_{\mathrm{max}}$ which maximizes the right hand side satisfies
\begin{eqnarray*}
U_{\mathrm{max}} \sqrt{\rho_0}E_?=\alpha \sqrt{\rho_1} E_?
\end{eqnarray*}
for some $\alpha \in \mathbb{C}$.
To find the unitary matrices $U_{\mathrm{max}}$ that maximize  $|\Tr(U \sqrt{\rho_0} \sqrt{\rho_1})|$  we use the following lemma:

\begin{lemma}
\label{maxUlemma}
For any operator $A$ in the space $M_n$ of $n\times n$ matrices we find
$$\max_U |\Tr(AU)|=Tr(|A|)$$
where the maximum is taken over all the unitary matrices.  The maximum is reached for any unitary operator $U$ that can be written as $U =V^{\dagger}e^{\imath \phi}$. Here $e^{\imath \phi}$ is an arbitrary phase while the unitary operator $V$ is defined via the polar decomposition $$A=|A| \, V$$ with $|A|=\sqrt{A A^\dagger}=V \, \sqrt{A^\dagger A}\, V^{\dagger}$.
\end{lemma}

\paragraph*{\bf Proof}
For any operator $A$, we can introduce its polar decomposition $A=|A| \, V$ with $|A|=\sqrt{A A^\dagger}=V \, \sqrt{A^\dagger A} \, V^\dagger$. Note that $V$ is unitary while $\sqrt{A A^\dagger}$ and $\sqrt{A^\dagger A}$ are unique, positive semi-definite and hermitian. With that we find
\begin{eqnarray*}
|\Tr(AU)|=|\Tr(|A|VU)|=|\Tr(|A|^{1/2}|A|^{1/2}VU)|.
\end{eqnarray*}
We denote $X=|A|^{1/2}=X^{\dagger}$ and $Y=|A|^{1/2}VU$ and apply the Cauchy-Schwarz inequality (Theorem 2) to obtain
\begin{eqnarray*}
|\Tr(AU)|=|\Tr(X^{\dagger}Y)| \le \sqrt{\Tr(|A|)} \,\, \sqrt{\Tr(U^{\dagger}V^{\dagger}|A|VU))}=\Tr(|A|) \; .
\end{eqnarray*}
Equality holds if and only if $|A|^{1/2}=\beta |A|^{1/2}VU$, for some $\beta \in {\mathbb C}$. This is possible if and only if $\beta V U= \openone$, where $U$ and $V$ are both unitary matrices. This means that $\beta=e^{-\imath \phi}$ for some $\phi$ so that we find the connection $U =V^{\dagger}e^{\imath \phi}$. This completes the proof of the lemma.

Thanks to lemma \ref{maxUlemma}, Eqn. (\ref{max}) implies
\begin{eqnarray*}
\sqrt{\Tr(E_? \rho_0)} \sqrt{\Tr(E_? \rho_1)}\ge |{\Tr(|\sqrt{\rho_0}\sqrt{\rho_1}|)|} \; 
\end{eqnarray*}
where equality now holds if and only if 
\begin{eqnarray} \label{condmax}
V^{\dagger}e^{\imath \phi} \sqrt{\rho_0}E_?=\alpha \sqrt{\rho_1} E_?
\end{eqnarray}
for some $\alpha \in \mathbb{C}$. Let us introduce the operators  $F_0:=|\sqrt{\rho_0}\sqrt{\rho_1}|=\sqrt{\sqrt{\rho_0}\rho_1\sqrt{\rho_0}}$ and $F_1=V^\dagger F_0 V=\sqrt{\sqrt{\rho_1}\rho_0\sqrt{\rho_1}}$, which are motivated by the polar decomposition
\begin{eqnarray}\label{PD}
\sqrt{\rho_0} \sqrt{\rho_1}= F_0 V=V F_1.
\end{eqnarray}
These operators are related to the fidelity of the two density matrices through the relation $F=\Tr(\sqrt{\sqrt{\rho_1}\rho_0\sqrt{\rho_1}})(=\Tr(F_0)=\Tr(F_1))$ (\cite{jozsa94a}). 

Next we use the definitions of the partial failure probabilities  $Q_i=\eta_i \Tr(E_? \rho_i)$ and  choose the phase $e^{\imath \phi}$ to be the same as the phase of $\alpha$ in (\ref{condmax})  to obtain the desired inequality $Q_0 Q_1\ge \eta_0 \eta_1 F^2$. Equality in the previous equation then holds if and only if $V^\dagger \sqrt{\rho_0}E_?=\alpha \sqrt{\rho_1} E_?$, for some $\alpha \in \mathbb{R}^+$. This completes the proof. \hfill $\blacksquare$

We can now derive the bounds in the different regimes of the ratio $\frac{\eta_1}{\eta_0}$ between the two {\it a priori} probabilities. Actually, the procedure is to find the minimum of the failure probability $Q=Q_0+Q_1$ under the constraints of the previous derived inequality $Q_0 Q_1\ge \eta_0 \eta_1 F^2$. According to Theorem 1, we can provide the necessary and sufficient condition for equality. 

\begin{theorem}
\label{3_bounds_theorem}
Let $\rho_0$ and $\rho_1$ be two density matrices with {\it a priori} probabilities $\eta_0$ and $\eta_1$. We define the fidelity $F$ of the two states $\rho_0$ and $\rho_1$ as $\Tr(\sqrt{\sqrt{\rho_0}\rho_1\sqrt{\rho_0}})$. We denote by $P_0$ and $P_1$, the projectors onto the support of $\rho_0$ and $\rho_1$. Then, for any USD measurement, the failure probability $Q$ obeys
\begin{eqnarray} \label{3_regimes}
Q \ge \eta_1 \frac{F^2}{\Tr(P_1 \rho_0)}+\eta_0 \Tr(P_1 \rho_0) \,\,\, &\textrm{for}& \,\,\, \sqrt{\frac{\eta_1}{\eta_0}} \le \frac{\Tr(P_1 \rho_0)}{F}\\ \nonumber
Q \ge 2\sqrt{\eta_0\eta_1}F \,\,\, &\textrm{for}& \,\,\, \frac{\Tr(P_1 \rho_0)}{F}\le \sqrt{\frac{\eta_1}{\eta_0}} \le \frac{F}{\Tr(P_0 \rho_1)}\\\nonumber
Q \ge \eta_0 \frac{F^2}{\Tr(P_0 \rho_1)}+\eta_1 \Tr(P_0 \rho_1) \,\,\, &\textrm{for}& \,\,\, \frac{F}{\Tr(P_0 \rho_1)} \le \sqrt{\frac{\eta_1}{\eta_0}} \; .\nonumber
\end{eqnarray}
Equality holds if and only if the unitary operator $V$ arising from a polar decomposition $\sqrt{\rho_0}\sqrt{\rho_1}=\sqrt{\sqrt{\rho_0}\rho_1\sqrt{\rho_0}}\,\, V$ satisfies  $V^\dagger \sqrt{\rho_0}E_?= \alpha \sqrt{\rho_1} E_?$, with $\alpha=\frac{\Tr(P_1 \rho_0)}{F}$, $\alpha =\sqrt{\frac{\eta_1}{\eta_0}}$  and $\alpha = \frac{F}{\Tr(P_0 \rho_1)}$ in the the first, second and third regime, respectively. 
\end{theorem}

\paragraph*{\bf Proof}
First of all, according to Theorem 1, we know that for any USD measurement the inequality $Q_1 \ge \frac{\eta_0 \eta_1F^2}{Q_0}$ holds. It follows that the failure probability is such that 
\begin{equation}
\label{optfunction}
Q \ge Q_0+\frac{\eta_0 \eta_1F^2}{Q_0}\; .
\end{equation}

Let us consider relations that only hold if  equality holds in Eqn. (\ref{optfunction}). In this case we have
\begin{equation}\label{Q1}
Q_0 Q_1 = \eta_0 \eta_1F^2 \;.
\end{equation}
Moreover, from Theorem 1 we know that in this case we have  $V^\dagger \sqrt{\rho_0}E_?= \alpha \sqrt{\rho_1} E_?$, for some $\alpha \in {\mathbb R}^+$. This relationship implies, via the respective definitions, that
\begin{equation}\label{Q2}
Q_0=\alpha^2 \frac{\eta_0}{\eta_1} Q_1 \; .
\end{equation}
We can combine the two equations (\ref{Q1}) and (\ref{Q2}) to
\begin{equation}
\label{alphadef}
Q_0=\alpha \eta_0 F \; .
\end{equation}
 So the final statement is that $Q=Q_0+\frac{\eta_0 \eta_1F^2}{Q_0}$ if and only if $V^\dagger \sqrt{\rho_0}E_?= \alpha \sqrt{\rho_1} E_?$, where $\alpha$ now is explicitly related to the other parameters as  $Q_0 = \alpha \eta_0 F$.

Second, we have to derive the range constraint on $Q_0$ and $Q_1$.  We know already that $Q_i \le \eta_i$. Moreover, from the work by Herzog and Bergou in \cite{herzog05a}, we learn that $\eta_0 \Tr(P_1 \rho_0) \le Q_0$ and $\eta_1 \Tr(P_0 \rho_1) \le Q_1$. Indeed, from the structure of the POVM elements, we have $E_0+E_1+E_?=\openone$ with $\mathcal{S}_{E_0} \subset \mathcal{S}^{\perp}_{\rho_1}$ and $\mathcal{S}_{E_1} \subset \mathcal{S}^{\perp}_{\rho_0}$. We consider only the non-trivial case where the supports of $\rho_0$ and $\rho_1$ are not identical. Then its structure must be such that $E_1+E_?=P_1 + R$ where $P_1$ is the projection onto the support of $\rho_1$ and $R$ is an hermitian positive semi-definite operator with support $\mathcal{S}_{R} \subset \mathcal{S}^\perp_{\rho_1}$ which satisfies $E_0+R = P_1^\perp$. Then it follows that $P_0=\eta_0 \Tr(E_0 \rho_0)=\eta_0 \Tr(P^\perp_1 \rho_0) - \eta_0 \Tr(R \rho_0)$. In our non-trivial case we will have $\Tr(R \rho_0)>0$ as soon as $R \neq 0$. This yields $P_0 \le \eta_0 \Tr(P^\perp_1 \rho_0)$ or equivalently $Q_0 \ge \eta_0 \Tr(P_1 \rho_0)$. In the same way, on can find $Q_1 \ge \eta_1 \Tr(P_0 \rho_1)$. We then have
\begin{eqnarray}\label{range}
\eta_0 \Tr(P_1 \rho_0) \le Q_0 \le \eta_0, \\
\eta_1 \Tr(P_0 \rho_1) \le Q_1 \le \eta_1. \nonumber
\end{eqnarray}
These two constraints can be combined to $\eta_0 \Tr(P_1 \rho_0) \le Q_0 \le \eta_0 \frac{F^2}{\Tr(P_0 \rho_1)}$. This can be seen as follows. Since $Q_1=\frac{\eta_0 \eta_1F^2}{Q_0}$, the constraints on $Q_1$ take the form  $\eta_0 F^2\le Q_0 \le \eta_0 \frac{F^2}{\Tr(P_0 \rho_1)}$. Let us consider the USD POVM given by $\{E_?=P_1,E_0=P_1^\perp,E_1=0\}$. Thank to Theorem 1, we find $\eta_0 \eta_1 F^2 \le \eta_0 \eta_1 \Tr(P_1 \rho_0) \Tr(P_1 \rho_1)$ or in other words $\eta_0 F^2 \le \eta_0 \Tr(P_1 \rho_0)$. We can also consider the USD POVM given by $\{E_?=P_0,E_0=0,E_1=P_0^\perp\}$ and with Theorem 1, we finally have $\eta_0\frac{F^2}{\Tr(P_0 \rho_1)} \le \eta_0$.

Next, we define the function $q(Q_0)=Q_0+\frac{\eta_0 \eta_1F^2}{Q_0}$  and minimize it under the constraint $\eta_0 \Tr(P_1 \rho_0) \le Q_0 \le \eta_0 \frac{F^2}{\Tr(P_0 \rho_1)}$. The resulting minimum will constitute a lower bound for $Q$. The function $q(Q_0)$ is convex ($\frac{d^2 q}{dQ_0^2}(Q_0) \ge 0$) and, therefore, it takes its minimum at the point $Q_0^{\text{min}}$ where the derivative vanishes ($\frac{d q}{dQ_0}(Q_0) = 0$ yielding $Q_0^{\text{min}}=\sqrt{\eta_0 \eta_1} F$) or at the limits of the constraint interval ($Q_0^{\text{min}}=\eta_0 \Tr(P_1 \rho_0)$ and $Q_0^{\text{min}}=\eta_0 \frac{F^2}{\Tr(P_0 \rho_1)}$). That gives us the minimum in three different regimes. In the first regime we have  $q_{\text{min}}(Q_0)=\eta_0 \Tr(P_1 \rho_0) + \eta_1 \frac{F^2}{\Tr(P_1 \rho_0)}$ and $Q_0^{\text{min}}=\eta_0 \Tr(P_1 \rho_0)$ if $\sqrt{\eta_0 \eta_1} F \le \eta_0 \Tr(P_1 \rho_0)$ that is to say if $\sqrt{\frac{\eta_1}{\eta_0}} \le \frac{\Tr(P_1 \rho_0)}{F}$. In the second regime we have  $q_{\text{min}}(Q_0)=2\sqrt{\eta_0\eta_1}F$ and $Q_0^{\text{min}}=\sqrt{\eta_0\eta_1}F$ if $\frac{\Tr(P_1 \rho_0)}{F} \le \sqrt{\frac{\eta_1}{\eta_0}} \le \frac{F}{\Tr(P_0 \rho_1)}$. The third regime gives  $q_{\text{min}}(Q_0)=\eta_0 \frac{F^2}{\Tr(P_0 \rho_1)}+\eta_1 \Tr(P_0 \rho_1)$ and $Q_0^{\text{min}}=\eta_0 \frac{F^2}{\Tr(P_0 \rho_1)}$ if $\frac{F}{\Tr(P_0 \rho_1)} \le \sqrt{\frac{\eta_1}{\eta_0}}$.

As a result we obtain lower bounds for the failure probability $Q$ in three regimes as given in Eqn. (\ref{3_regimes}). For each regime, the value of $Q_0$ which minimized $q(Q_0)$ is given and via Eqn. (\ref{alphadef}) we find the corresponding value that $\alpha$ has to take. We read off the values as  $\alpha=\frac{\Tr(P_1 \rho_0)}{F}$, $\alpha=\sqrt{\frac{\eta_1}{\eta_0}}$ and $\alpha=\frac{F}{\Tr(P_0 \rho_1)}$ for the first, second and third regime, respectively. \hfill $\blacksquare$

Let us note that, by construction, those bounds are tighter than the ones in \cite{rudolph03a}. Indeed, one could recover the three bounds in \cite{rudolph03a} by looking for the minimum of the function $q(Q_0)$ under the weaker constraints $\eta_0 F^2 \le Q_0 \le \eta_0$.

\section{The parallel addition $\rho_0:\rho_1$}
\label{sec:parallel}

Before deriving our central theorem, we will first recall some useful results of linear algebra. We denote by $M^{-1}$ the pseudo-inverse of a matrix $M$, which has not necessarily full rank. The pseudo-inverse can be defined via the singular-value decomposition of $M$. Whenever $M$ is of full rank, the pseudo-inverse coincides with the inverse.  In general, it is not known how to express the pseudo inverse of a sum $(A+B)^{-1}$  in terms of the pseudo inverses $A^{-1}$ and $B^{-1}$ \cite{fill98a}. However, a related new operation $A(A+B)^{-1}B$, called parallel addition and denoted by $A:B$ has been defined in 1969 by Anderson and Duffin and will turn out useful in our context. 

First of all, we denote by ${\mathcal S}_M$, the support of a hermitian and positive semi-definite matrix $M$. We then have the following property for the parallel addition:

\begin{property}\label{parallel}
{\rm\cite{anderson69a}} Let $A$ and $B$ be two hermitian and positive semi-definite matrices in $M_n$, then the support ${\mathcal S}_{A:B}$ of $A:B$ is given in terms of the supports of $A$ and $B$ as
$${\mathcal S}_{A:B}={\mathcal S}_A \cap {\mathcal S}_B.$$
\end{property}

Next let us recall two reduction theorem for USD of mixed states \cite{raynal03a}.  We consider the problem of discriminating unambiguously two density matrices $\rho_0$ and $\rho_1$ with {\it a priori} probabilities $\eta_0$ and $\eta_1$. We denote by $r_0$ the rank of $\rho_0$ and by $r_1$ the rank of $\rho_1$. A general USD problem can satisfy $r_0+r_1 \ge d$, where $d$ is the dimension of the Hilbert space ${\mathcal H}$ spanned by the two states. This means in particular that the two supports can overlap.

 In a first reduction theorem it has been  shown by the authors \cite{raynal03a} that any such USD problem can always be reduced to the one of discriminating $\rho'_0$ and $\rho'_1$, two density matrices of rank $r'_0$ and $r'_1$ with {\it a priori} probabilities $\eta'_0$ and $\eta'_1$, spanning the same Hilbert space ${\mathcal H}$ of dimension $r'_0+r'_1$. Indeed we can split off any common subspace of the supports ${\mathcal S}_{\rho_0} \cap {\mathcal S}_{\rho_1}$ to end up with ${\mathcal S}_{\rho'_0} \cap {\mathcal S}_{\rho'_1}=\{0\}$. An easy way to know whether the two supports overlap is to check whether the equality $rk(\rho'_0)+rk(\rho'_1)=rk(\rho'_0 + \rho'_1)$ holds (see details in \cite{marsaglia72a}). In the reduced case, property (\ref{parallel}) implies ${\mathcal S}_{\rho'_0:\rho'_1}=0$ that is to say $\rho'_0:\rho'_1=0$. By defining $\Sigma:=\rho'_0+\rho'_1$, we can write the parallel addition as  $\rho'_0 \Sigma^{-1} \rho'_1$. Moreover, since $rk(\rho'_0 + \rho'_1)=\text{dim}({\mathcal H})$, we end up with $\Sigma$ having full-rank and $\Sigma \Sigma^{-1}=\openone_{\mathcal H}$.

We therefore  have the following corollary to property (1),
\begin{corollary}
Let $\rho_0$ and $\rho_1$ be two density matrices spanning a Hilbert space ${\mathcal H}$. Let $\Sigma$ be defined as the sum of these two density matrices. $${\rm If}\,\,\, rk(\rho_0)+rk(\rho_1)=rk(\rho_0 + \rho_1)\,\,\, {\rm then}\,\,\, \rho_0\Sigma^{-1}\rho_1=0.$$
\end{corollary}

According to the first reduction theorem we can, without loss of generality, consider only USD problems of two density matrices without overlap of their supports. In the following, we consider two density matrices $\rho_0$ and $\rho_1$ (which are hermitian and positive semi-definite matrices) such that $rk(\rho_0 + \rho_1)=rk(\rho_0)+rk(\rho_1)=\text{dim}({\mathcal H})$. As explained above, for such a problem, $\rho_0\Sigma^{-1}\rho_1=0$, with $\Sigma = \rho_0 + \rho_1$ having full rank. This leads to $\rho_0\Sigma^{-1}\rho_0=\rho_0$ and $\rho_1\Sigma^{-1}\rho_1=\rho_1$ since $\Sigma \Sigma^{-1}=\openone_{\mathcal H}$. The projectors onto the supports of those two density matrices can then be written as : $P_{\rho_0}=\sqrt{\rho_0}\Sigma^{-1}\sqrt{\rho_0}$ and $P_{\rho_1}=\sqrt{\rho_1}\Sigma^{-1}\sqrt{\rho_1}$.

A second reduction theorem \cite{raynal03a} allows to eliminate the part of the support of $\rho'_0$ which is orthogonal to the support of $\rho'_1$ and {\it vice et versa}. This implies that the two resulting density matrices $\rho''_0$ and $\rho''_1$ possess the same rank $r$ and span a $2r$-dimensional Hilbert space. We denote such a USD problem by "$r+r=2r$" (see \cite{raynal03a} for more details). This second reduction theorem indicates in which situations a further reduction of the original problem can be achieved. This theorem is not needed for the derivation of our central theorem.

\section{Necessary and sufficient conditions}
\label{conditions}

We are now ready to derive the main result of this paper. The first part of this result gives compact necessary and sufficient conditions for a pair of mixed states to saturate the bounds of the failure probability $Q$. The second part gives  the corresponding POVMs in an explicit form.  To clarify the notation, let us note that in  $M \ge 0$ we mean that the  operator $M$ is hermitian and positive semi-definite.

\begin{theorem}
\label{boundsaturation}
Necessary and sufficient conditions to saturate the bounds on the failure probability\\
Consider a USD problem defined by the two density matrices $\rho_0$ and $\rho_1$ and their respective {\it a priori} probabilities $\eta_0$ and $\eta_1$ such that their supports satisfy ${\mathcal S}_{\rho_0} \cap {\mathcal S}_{\rho_1}=\{0\}$ (Any USD problem of two density matrices can be reduced to such a form according to \cite{raynal03a}). Let $F_0$ and $F_1$ be the two operators $\sqrt{\sqrt{\rho_0}\rho_1\sqrt{\rho_0}}$ and $\sqrt{\sqrt{\rho_1}\rho_0\sqrt{\rho_1}}$. The fidelity $F$ of the two states $\rho_0$ and $\rho_1$ is then given by $\Tr(F_0)=\Tr(F_1)$. We denote by $P_0$ and $P_1$, the projectors onto the support of $\rho_0$ and $\rho_1$. The optimal failure probability $Q^{\textrm{opt}}$ for USD then satisfies 

\begin{eqnarray} \label{theocentral}
Q^{\mathrm{opt}} = \eta_1 \frac{F^2}{\Tr(P_1 \rho_0)}+\eta_0 \Tr(P_1 \rho_0) \, & \Leftrightarrow & \,
\begin{array}{c}
\rho_0-\frac{\Tr(P_1 \rho_0)}{F} F_0 \ge 0 \\ 
\rho_1-\frac{F}{\Tr(P_1 \rho_0)}F_1 \ge 0 \\ 
\end{array} \,\,\, \mathrm{for} \,\,\, \sqrt{\frac{\eta_1}{\eta_0}} \le \frac{\Tr(P_1 \rho_0)}{F}\\ \nonumber
\\
Q^{\mathrm{opt}} = 2\sqrt{\eta_0\eta_1}F  & \Leftrightarrow & \, 
\begin{array}{cc}
\rho_0-\sqrt{\frac{\eta_1}{\eta_0}}F_0 \ge 0 \\ 
\rho_1-\sqrt{\frac{\eta_0}{\eta_1}}F_1 \ge 0 \\ 
\end{array} \,\,\, \mathrm{for} \,\,\, \frac{\Tr(P_1 \rho_0)}{F}\le \sqrt{\frac{\eta_1}{\eta_0}} \le \frac{F}{\Tr(P_0 \rho_1)}\\ \nonumber
\\
Q^{\mathrm{opt}} = \eta_0 \frac{F^2}{\Tr(P_0 \rho_1)}+\eta_1 \Tr(P_0 \rho_1) & \Leftrightarrow & \, 
\begin{array}{cc}
\rho_0-\frac{F}{\Tr(P_0 \rho_1)}F_0 \ge 0 \\ 
\rho_1-\frac{\Tr(P_0 \rho_1)}{F} F_1 \ge 0 \\ 
\end{array} \,\,\, \mathrm{for} \,\,\, \frac{F}{\Tr(P_0 \rho_1)} \le \sqrt{\frac{\eta_1}{\eta_0}} \nonumber
\end{eqnarray}

The POVM elements that realize these optimal failure probabilities, if the corresponding conditions are fulfilled,  are given by
\begin{eqnarray}
E_0&=&\Sigma^{-1} \sqrt{\rho_0} \left(\rho_0-\alpha F_0 \right) \sqrt{\rho_0}\Sigma^{-1} \\ \nonumber
E_1&=&\Sigma^{-1} \sqrt{\rho_1} \left(\rho_1-\frac{1}{\alpha} F_1\right)\sqrt{\rho_1}\Sigma^{-1} \\ \nonumber
E_?&=&\Sigma^{-1} \left(\sqrt{\alpha} \sqrt{\rho_0}+\frac{1}{\sqrt{\alpha}} \sqrt{\rho_1} V^\dagger\right) F_0 \left(\sqrt{\alpha} \sqrt{\rho_0}+\frac{1}{\sqrt{\alpha}}V \sqrt{\rho_1}\right)\Sigma^{-1} 
\end{eqnarray}
with $\alpha=\frac{\Tr(P_1 \rho_0)}{F}$ for the first regime, $\alpha=\sqrt{\frac{\eta_1}{\eta_0}}$ for the second regime and $\alpha=\frac{F}{\Tr(P_0 \rho_1)}$ for the third regime.

\end{theorem}

\paragraph*{\bf Proof}

First, we give a proof for the necessary conditions.\\

\paragraph*{ Proof for the necessary conditions}

From Theorem \ref{3_bounds_theorem} we know that the bounds on the failure probability are satisfied whenever  $V^\dagger \sqrt{\rho_0}E_?=\alpha \sqrt{\rho_1} E_?$ with $\alpha=\frac{\Tr(P_1 \rho_0)}{F}$, $\alpha=\sqrt{\frac{\eta_1}{\eta_0}}$ and $\alpha=\frac{F}{\Tr(P_0 \rho_1)}$ for the three regimes. 

We replace $E_?$ by $\openone-E_0-E_1$, multiply on the left by $V$ and on the right by $\sqrt{\rho_0}$. This leads us to
\begin{eqnarray}\label{pos}
\rho_0-\alpha F_0=\sqrt{\rho_0}E_0\sqrt{\rho_0}
\end{eqnarray}
where we used the relation (\ref{PD}) $\sqrt{\rho_0} \sqrt{\rho_1}= F_0 V$ and the fact that the support of $\rho_i$ and $E_j$ are orthogonal for $i\neq j$. Indeed, let us notice that $\Tr(E_i\rho_j)=0 \Leftrightarrow E_i\rho_j=0$ because $E_i$ and $\rho_j$ are hermitian and positive semi-definite operators \cite{raynal03a}. The right hand side in (\ref{pos}) is hermitian and positive semi-definite because of the form $AA^\dagger$ with $A=\sqrt{\rho_0}\sqrt{E_0}$. Then $\rho_0-\alpha F_0$ must be hermitian and positive semi-definite as well. A similar calculation where we only multiply on the right by $\sqrt{\rho_1}$ instead of by $\sqrt{\rho_0}$ leads us to $$\rho_1-\frac{1}{\alpha} F_1=\sqrt{\rho_1}E_1\sqrt{\rho_1}$$ which is again a hermitian and positive semi-definite operator.\\

With that we have proved that if equality holds in the bounds of Theorem \ref{3_bounds_theorem} then we have
\begin{eqnarray}
\label{posop}
\rho_0-\alpha F_0 &\geq& 0\\
\rho_1-\frac{1}{\alpha} F_1&\geq& 0 \; ,\nonumber
\end{eqnarray}
which form, therefore, necessary conditions for equality in the bounds of Theorem \ref{3_bounds_theorem}. 

\paragraph*{ Proof for the sufficient conditions}
Now we start with the assumption that the conditions (\ref{posop}) are fulfilled. 
Let us define the following POVM elements : 
\begin{eqnarray} \label{POVM}
E_0&=&\Sigma^{-1} \sqrt{\rho_0} \left(\rho_0-\alpha F_0 \right)\sqrt{\rho_0}\Sigma^{-1} \\ \nonumber
E_1&=&\Sigma^{-1} \sqrt{\rho_1} \left(\rho_1-\frac{1}{\alpha} F_1 \right)\sqrt{\rho_1}\Sigma^{-1} \\ \nonumber
E_?&=&\Sigma^{-1} \left(\sqrt{\alpha} \sqrt{\rho_0}+\frac{1}{\sqrt{\alpha}} \sqrt{\rho_1} V^{\dagger} \right) F_0 \left(\sqrt{\alpha} \sqrt{\rho_0}+\frac{1}{\sqrt{\alpha}}V \sqrt{\rho_1}\right)\Sigma^{-1} \nonumber
\end{eqnarray}

First, let us verify that this is indeed a valid POVM. The three operators are positive since they are of the form $A\dagger M A$ where $M$ is a positive hermitian operator. In the first two cases this is true because of the conditions (\ref{posop}), in the third case it follows from the positivity of $F_0$. The three operators sum up to identity,  $E_0+E_1+E_? = \openone$, as can be checked by straight forward calculation which makes use also of Eqn. (\ref{PD}).  
Next, we have to check that the given POVM is a valid USD POVM, that is, $\Tr(\rho_0 E_1)=\Tr(\rho_1 E_0)=0$. This relation holds since the supports of $\rho_0$ and $\rho_1$ do not  overlap. Therefore,  corollary 1 applies and we have  $\rho_0 \Sigma^{-1} \rho_1=0$ from which follows that   $\sqrt{\rho_0} \Sigma^{-1} \rho_1=0$ and $\sqrt{\rho_1} \Sigma^{-1} \rho_0=0$.Finally, one can check in a straight forward calculation exploiting the properties used in the previous checks that this  POVM lead to the three desired failure probabilities.

 Let us note that we have only used the assumption about the non-overlapping supports to prove the sufficiency of the conditions. Their necessity does not require this assumption. \hfill $\blacksquare$

\section{Discussion}
\label{discussion}
Theorem \ref{boundsaturation} characterizes under which circumstances the equality of the bounds in Theorem \ref{3_bounds_theorem} can be obtained. Whenever two mixed density matrices have no overlapping supports and the corresponding two operators in Theorem \ref{boundsaturation} are positive semidefinite, we can give explicitly the optimum USD POVM.

The first question is to know whether the set  of pairs of generic mixed states (a USD problem which is not reducible to some pure state case), that fulfill the constraints $\rho_0-\sqrt{\frac{\eta_1}{\eta_0}} F_0 \ge 0$ and $\rho_1-\sqrt{\frac{\eta_0}{\eta_1}} F_1 \ge 0$, is empty or not. Actually this set is non-empty. For instance, consider a problem  motivated by a four-state quantum key distribution protocol using  coherent states \cite{dusek00a}. Here it might be of interest for an eavesdropper to distinguish the density matrices  $\rho_0=\frac{1}{2}[|\alpha\rangle \langle \alpha|+|-\alpha\rangle \langle -\alpha|]$ and $\rho_1=\frac{1}{2}[|\imath\alpha\rangle \langle \imath \alpha|+|-\imath \alpha\rangle \langle -\imath \alpha|]$, corresponding to the bit value $0$ and $1$, respectively.   In fact, this pair of states can be represented as geometrically uniform (GU) states \cite{eldar01a} as they are related as $\rho_1=U\rho_0U^{\dagger}$ with $U^2=\openone$. They can be represented as operators over a four-dimensional Hilbert space as 
\begin{eqnarray}
\rho_0=\left(
\begin{array}{cccc}
|c_0|^2& 0& c_0c_2*& 0\\
0& |c_1|^2& 0& c_1c_3*\\
c_2c_0*& 0& |c_2|^2& 0\\
0& c_3c_1*& 0& |c_3|^2
\end{array}
\right)
\end{eqnarray}
with complex coefficients $c_i$ depending on phase and as given in \cite{dusek00a}, and
\begin{eqnarray}
U=\left(
\begin{array}{cccc}
-1 & 0 & 0 & 0\\
0 & -1 & 0 & 0\\
0 & 0 & 1 & 0\\
0 & 0 & 0 & 1 \end{array}
\right)\;.
\end{eqnarray}
One can show that for these two states the operators $\rho_0-\sqrt{\frac{\eta_1}{\eta_0}}F_0$ and $\rho_1-\sqrt{\frac{\eta_0}{\eta_1}}F_1$ are hermitian and positive semi-definite for some regime of the ratio $\frac{\eta_1}{\eta_0}$ around the value $\frac{\eta_1}{\eta_0}=1$ included into the second regime (for any $c_0$, $c_1$, $c_2$ and $c_3$ in ${\mathbb C}$). According to Theorem \ref{boundsaturation}, the optimal failure probability is $Q^{\mathrm{opt}} = 2\sqrt{\eta_0\eta_1}F$ where the fidelity is given by $F=e^{\frac{-|\alpha|^2}{2}}(|\cos{\frac{|\alpha|^2}{2}}|+|\sin{\frac{|\alpha|^2}{2}}|)$. Let us note that, in general, those operators are not positive for the whole second regime $F \leq  \sqrt{\frac{\eta_1}{\eta_0}} \leq \frac{1}{F}$. Actually this depends on the parameters $c_0$, $c_1$, $c_2$ and $c_3$ in ${\mathbb C}$. This implies that, in general, the necessary and sufficient conditions are not fulfilled neither for the first regime nor for the third regime for these two coherent states.

Actually two GU states are not necessarily in the set of states that saturate the bound, not even for equal {\em a priori} probabilities. For example, one can consider the two GU states $\rho_0$ and $\rho'_1=W\rho_0W^{\dagger}$ where $\rho_0$ is given as above while 
\begin{eqnarray}
W=\frac{1}{\sqrt{2}}\left(
\begin{array}{cccc}
1 & 1 & 0 & 0\\
1 & -1 & 0 & 0\\
0 & 0 & 1 & 1\\
0 & 0 & 1 & -1 \end{array}
\right)
\end{eqnarray}
with $c_0=\sqrt{0.1}$, $c_1=\sqrt{0.4}$, $c_2=\sqrt{0.3}$ and $c_3=\sqrt{0.2}$ and $\eta_0=\eta_1$. Those states are indeed GU states since $W^2=\openone$. The supports do not overlap. However,   one can show that the operators $\rho_0-F_0$ and $\rho_1-F_1$ are not positive semi-definite.

As a result, there exist generic mixed states that satisfy the conditions of Theorem 4 and for which a optimal USD strategy can be given. However, there are generic mixed states that do not satisfy the conditions so that  it remains to find the optimal failure probability in those cases.\\

The second remark is about the link between our result and the pure state case. Actually for two pure states, since  $F_0=F|\Psi_0\rangle \langle \Psi_0|$, $F_1=F|\Psi_1\rangle \langle \Psi_1|$ and $\Tr(P_0\rho_1)=\Tr(P_1 \rho_0)=F^2$, the constraints $\rho_0-\alpha F_0 \ge 0$, $\rho_1-\frac{1}{\alpha} F_1 \ge 0$ are always fulfilled and our result reduces to the one of Shimony and Jaeger. We can go beyond this remark and find under which conditions our bounds reduce to the ones in \cite{rudolph03a}. Since our bounds are tighter, the bounds in \cite{rudolph03a} are reached if and only if, first, the condition in Theorem 4 are fulfilled and, second, the equalities $\Tr(P_0\rho_1)=\Tr(P_1 \rho_0)=F^2$ hold (like in the pure state case). This is made more precise in the following corollary to Theorem 4:

\begin{corollary}
Necessary and sufficient conditions to saturate the bounds in \cite{rudolph03a}\\
Consider a USD problem defined by the two density matrices $\rho_0$ and $\rho_1$ and their respective {\it a priori} probabilities $\eta_0$ and $\eta_1$ such that their supports satisfy ${\mathcal S}_{\rho_0} \cap {\mathcal S}_{\rho_1}=\{0\}$ (Any USD problem of two density matrices can be reduced to such a form according to \cite{raynal03a}). Let $F_0$ and $F_1$ be the two operators $\sqrt{\sqrt{\rho_0}\rho_1\sqrt{\rho_0}}$ and $\sqrt{\sqrt{\rho_1}\rho_0\sqrt{\rho_1}}$. The fidelity $F$ of the two states $\rho_0$ and $\rho_1$ is then given by $\Tr(F_0)=\Tr(F_1)$. We denote by $P_0$ and $P_1$, the projectors onto the support of $\rho_0$ and $\rho_1$. The optimal failure probability $Q^{\textrm{opt}}$ for USD then satisfies

\begin{eqnarray} \label{theocentral}
Q^{\mathrm{opt}} = \eta_1+\eta_0 F^2 \, & \Leftrightarrow & \,
\begin{array}{c}
\rho_0-F F_0 \ge 0 \\ 
\rho_1-\frac{1}{F}F_1 = 0 \\ 
\end{array} \,\,\, \mathrm{for} \,\,\, \sqrt{\frac{\eta_1}{\eta_0}} \le F\\ \nonumber
\\
Q^{\mathrm{opt}} = 2\sqrt{\eta_0\eta_1}F  & \Leftrightarrow & \, 
\begin{array}{cc}
\rho_0-\sqrt{\frac{\eta_1}{\eta_0}}F_0 \ge 0 \\ 
\rho_1-\sqrt{\frac{\eta_0}{\eta_1}}F_1 \ge 0 \\ 
\end{array} \,\,\, \mathrm{for} \,\,\, F\le \sqrt{\frac{\eta_1}{\eta_0}} \le \frac{1}{F}\\ \nonumber
\\
Q^{\mathrm{opt}} = \eta_0+\eta_1 F^2 & \Leftrightarrow & \, 
\begin{array}{cc}
\rho_0-\frac{1}{F}F_0 = 0 \\ 
\rho_1-F F_1 \ge 0 \\ 
\end{array} \,\,\, \mathrm{for} \,\,\, \frac{1}{F} \le \sqrt{\frac{\eta_1}{\eta_0}} \nonumber
\end{eqnarray}

The POVM elements that realize these optimal failure probabilities, if the corresponding conditions are fulfilled,  are given by
\begin{eqnarray}
E_0&=&\Sigma^{-1} \sqrt{\rho_0} \left(\rho_0-\alpha F_0 \right) \sqrt{\rho_0}\Sigma^{-1} \\ \nonumber
E_1&=&\Sigma^{-1} \sqrt{\rho_1} \left(\rho_1-\frac{1}{\alpha} F_1\right)\sqrt{\rho_1}\Sigma^{-1} \\ \nonumber
E_?&=&\Sigma^{-1} \left(\sqrt{\alpha} \sqrt{\rho_0}+\frac{1}{\sqrt{\alpha}} \sqrt{\rho_1} V^\dagger\right) F_0 \left(\sqrt{\alpha} \sqrt{\rho_0}+\frac{1}{\sqrt{\alpha}}V \sqrt{\rho_1}\right)\Sigma^{-1} 
\end{eqnarray}
with $\alpha=F$ for the first regime, $\alpha=\sqrt{\frac{\eta_1}{\eta_0}}$ for the second regime and $\alpha=\frac{1}{F}$ for the third regime.
\end{corollary}

In the first regime we find that  $E_1 = 0$ because this operator is hermitian, positive semi-definite and its trace vanishes. The resulting POVM has to be a projective measurement with projections onto the support of $\rho_1$ and onto its orthogonal complement, i.e.\ $E_0=P^\perp_1$, $E_1=0$ and $E_?=P_1$. A direct proof from the explicit expressions in Eqn. (\ref{POVM}) is difficult, however a simple reasoning allows to verify this statement. We consider only the non-trivial case where the supports of $\rho_1$ and $\rho_2$ are not identical. Of course, a two-element USD POVM satisfies $E_0+E_?=\openone$ with $\mathcal{S}_{E_0} \subset \mathcal{S}_{\rho_1}$. Then its structure must be such that $E_?=P_1 + R$ where $P_1$ is the projection onto the support of $\rho_1$ and $R$ is an operator with support $\mathcal{S}_{R} \subset \mathcal{S}^\perp_{\rho_1}$ which satisfies $E_0+R = P_1^\perp$. Then it follows that $Q=\eta_1 + \eta_0 \Tr(P_1 \rho_0) + \eta_0 \Tr(R \rho_0)$. In our non-trivial case we will have $\Tr(R \rho_0)>0$ as soon as $R \neq 0$. Therefore we find as an optimal solution within this class of two-element USD POVM, the POVM with $R=0$ leading to $E_?=P_1$ and $E_0=P_1^\perp$. We can actually write the failure probability as $Q^{\mathrm{opt}} = \eta_1+\eta_0 F^2$. Indeed $\rho_1=\frac{1}{F}F_1$ then $\rho^2_1=\frac{1}{F^2} \sqrt{\rho_1}\rho_0 \sqrt{\rho_1}$. This implies $F^2\rho_1=P_1\rho_0P_1$ and finally $\Tr(P_1 \rho_0)=F^2$. This is consistent with the results derived above and gives 
the correct failure probability.
In the third regime, we have $E_0=0$ and the corresponding POVM is a projective measurement with $E_0=0$, $E_1=P^\perp_0$, $E_?=P_0$.

Finally, let us note that the optimal error probability for the minimum error discrimination strategy is $Q^{\mathrm{opt}}_{\mathrm{MED}}=\frac{1}{2}\left( 1-\Tr(|\eta_1 \rho_1-\eta_0 \rho_0|\right)$  \cite{helstrom76a}. Then, on one hand the trace distance is related to the minimum error discrimination while on the other hand the Fidelity is related to the unambiguous state discrimination strategy.

\section{Conclusion}
\label{conclusions}

To summarize, we have given new bounds on the failure probability of unambiguously discriminating two mixed states. Moreover, we provide necessary and sufficient conditions for two mixed states to saturate those bounds. With that result, we give the optimal USD POVM of a wide class of pairs of mixed states. This class corresponds to pairs of mixed states for which the lower bounds (one for each of the three regimes depending on the ratio between the {\it a priori} probabilities) on the failure probability $Q$ are saturated. This class in non empty since it contains some pairs of generic mixed states as well as any pair of pure states. For those pairs, we provide the first analytical solutions for unambiguous discrimination of generic mixed states. This goes beyond known results which are all reducible to some pure state case. Additionally, we showed that there exists pairs of mixed states that cannot saturate the bounds.

\begin{center}
{\bf Acknowledgments}
\end{center}

We would like to thank Janos Bergou for discussions and drawing our attention to the problem whether the bounds found by Rudolph {\it et al.} can always be reached. Further, we thank Ulrike Herzog for transmitting a manuscript of \cite{herzog05a} prior for publication. Finally we thank Aska Dolinska and the whole QIT group for very useful discussions. This work was supported by the DFG under the Emmy-Noether program, the EU FET network RAMBOQ (IST-2002-6.2.1) and the network of competence QIP of the state of Bavaria (A8).

\bibliography{qit_neu}

\begin{thebibliography}{23}
\expandafter\ifx\csname natexlab\endcsname\relax\def\natexlab#1{#1}\fi
\expandafter\ifx\csname bibnamefont\endcsname\relax
  \def\bibnamefont#1{#1}\fi
\expandafter\ifx\csname bibfnamefont\endcsname\relax
  \def\bibfnamefont#1{#1}\fi
\expandafter\ifx\csname citenamefont\endcsname\relax
  \def\citenamefont#1{#1}\fi
\expandafter\ifx\csname url\endcsname\relax
  \def\url#1{\texttt{#1}}\fi
\expandafter\ifx\csname urlprefix\endcsname\relax\def\urlprefix{URL }\fi
\providecommand{\bibinfo}[2]{#2}
\providecommand{\eprint}[2][]{\url{#2}}

\bibitem[{\citenamefont{Rudolph et~al.}(2003)\citenamefont{Rudolph, Spekkens,
  and Turner}}]{rudolph03a}
\bibinfo{author}{\bibfnamefont{T.}~\bibnamefont{Rudolph}},
  \bibinfo{author}{\bibfnamefont{R.~W.} \bibnamefont{Spekkens}},
  \bibnamefont{and} \bibinfo{author}{\bibfnamefont{P.~S.}
  \bibnamefont{Turner}}, \bibinfo{journal}{Phys. Rev. A}
  \textbf{\bibinfo{volume}{68}}, \bibinfo{pages}{010301(R)}
  (\bibinfo{year}{2003}).

\bibitem[{\citenamefont{Helstrom}(1976)}]{helstrom76a}
\bibinfo{author}{\bibfnamefont{C.~W.} \bibnamefont{Helstrom}},
  \emph{\bibinfo{title}{Quantum detection and estimation theory}}
  (\bibinfo{publisher}{Academic Press}, \bibinfo{address}{New York},
  \bibinfo{year}{1976}).

\bibitem[{\citenamefont{Dieks}(1988)}]{dieks88a}
\bibinfo{author}{\bibfnamefont{D.}~\bibnamefont{Dieks}},
  \bibinfo{journal}{Phys. Lett. A} \textbf{\bibinfo{volume}{126}},
  \bibinfo{pages}{303} (\bibinfo{year}{1988}).

\bibitem[{\citenamefont{Ivanovic}(1987)}]{ivanovic87a}
\bibinfo{author}{\bibfnamefont{I.~D.} \bibnamefont{Ivanovic}},
  \bibinfo{journal}{Phys. Lett. A} \textbf{\bibinfo{volume}{123}},
  \bibinfo{pages}{257} (\bibinfo{year}{1987}).

\bibitem[{\citenamefont{Peres}(1988)}]{peres88a}
\bibinfo{author}{\bibfnamefont{A.}~\bibnamefont{Peres}},
  \bibinfo{journal}{Phys. Lett. A} \textbf{\bibinfo{volume}{128}},
  \bibinfo{pages}{19} (\bibinfo{year}{1988}).

\bibitem[{\citenamefont{Peres and Terno}(1998)}]{peres98a}
\bibinfo{author}{\bibfnamefont{A.}~\bibnamefont{Peres}} \bibnamefont{and}
  \bibinfo{author}{\bibfnamefont{D.~R.} \bibnamefont{Terno}},
  \bibinfo{journal}{J. Phys. A:Math. Gen.} \textbf{\bibinfo{volume}{31}},
  \bibinfo{pages}{7105} (\bibinfo{year}{1998}).

\bibitem[{\citenamefont{Chefles and Barnett}(1998)}]{chefles98a}
\bibinfo{author}{\bibfnamefont{A.}~\bibnamefont{Chefles}} \bibnamefont{and}
  \bibinfo{author}{\bibfnamefont{S.~M.} \bibnamefont{Barnett}},
  \bibinfo{journal}{Phys. Lett. A} \textbf{\bibinfo{volume}{250}},
  \bibinfo{pages}{223} (\bibinfo{year}{1998}).

\bibitem[{\citenamefont{Jaeger and Shimony}(1995)}]{jaeger95a}
\bibinfo{author}{\bibfnamefont{G.}~\bibnamefont{Jaeger}} \bibnamefont{and}
  \bibinfo{author}{\bibfnamefont{A.}~\bibnamefont{Shimony}},
  \bibinfo{journal}{Phys. Lett. A} \textbf{\bibinfo{volume}{197}},
  \bibinfo{pages}{83} (\bibinfo{year}{1995}).

\bibitem[{\citenamefont{Sun et~al.}(2002)\citenamefont{Sun, Bergou, and
  Hillery}}]{sun02a}
\bibinfo{author}{\bibfnamefont{Y.}~\bibnamefont{Sun}},
  \bibinfo{author}{\bibfnamefont{J.~A.} \bibnamefont{Bergou}},
  \bibnamefont{and} \bibinfo{author}{\bibfnamefont{M.}~\bibnamefont{Hillery}},
  \bibinfo{journal}{Phys. Rev. A} \textbf{\bibinfo{volume}{66}},
  \bibinfo{pages}{032315} (\bibinfo{year}{2002}).

\bibitem[{\citenamefont{Bergou et~al.}(2003)\citenamefont{Bergou, Herzog, and
  Hillery}}]{bergou03a}
\bibinfo{author}{\bibfnamefont{J.~A.} \bibnamefont{Bergou}},
  \bibinfo{author}{\bibfnamefont{U.}~\bibnamefont{Herzog}}, \bibnamefont{and}
  \bibinfo{author}{\bibfnamefont{M.}~\bibnamefont{Hillery}},
  \bibinfo{journal}{Phys. Rev. Lett.} \textbf{\bibinfo{volume}{90}},
  \bibinfo{pages}{257901} (\bibinfo{year}{2003}).

\bibitem[{\citenamefont{Herzog and Bergou}(2005)}]{herzog05a}
\bibinfo{author}{\bibfnamefont{U.}~\bibnamefont{Herzog}} \bibnamefont{and}
  \bibinfo{author}{\bibfnamefont{J.}~\bibnamefont{Bergou}},
  \bibinfo{journal}{quant-ph/0502117}  (\bibinfo{year}{2005}).

\bibitem[{\citenamefont{Raynal et~al.}(2003)\citenamefont{Raynal, L\"utkenhaus,
  and van Enk}}]{raynal03a}
\bibinfo{author}{\bibfnamefont{P.}~\bibnamefont{Raynal}},
  \bibinfo{author}{\bibfnamefont{N.}~\bibnamefont{L\"utkenhaus}},
  \bibnamefont{and} \bibinfo{author}{\bibfnamefont{S.}~\bibnamefont{van Enk}},
  \bibinfo{journal}{Phys. Rev. A} \textbf{\bibinfo{volume}{68}},
  \bibinfo{pages}{022308} (\bibinfo{year}{2003}).

\bibitem[{\citenamefont{Fiurasek and Jezek}(2003)}]{fiurasek03a}
\bibinfo{author}{\bibfnamefont{J.}~\bibnamefont{Fiurasek}} \bibnamefont{and}
  \bibinfo{author}{\bibfnamefont{M.}~\bibnamefont{Jezek}},
  \bibinfo{journal}{Phys. Rev. A} \textbf{\bibinfo{volume}{67}},
  \bibinfo{pages}{012321} (\bibinfo{year}{2003}).

\bibitem[{\citenamefont{Eldar et~al.}(2004)\citenamefont{Eldar, Stojnic, and
  Hassabi}}]{eldar04a}
\bibinfo{author}{\bibfnamefont{Y.~C.} \bibnamefont{Eldar}},
  \bibinfo{author}{\bibfnamefont{M.}~\bibnamefont{Stojnic}}, \bibnamefont{and}
  \bibinfo{author}{\bibfnamefont{B.}~\bibnamefont{Hassabi}},
  \bibinfo{journal}{Phys. Rev. A} \textbf{\bibinfo{volume}{69}},
  \bibinfo{pages}{062318} (\bibinfo{year}{2004}).

\bibitem[{\citenamefont{Feng et~al.}(2004)\citenamefont{Feng, Duan, and
  Ying}}]{feng04a}
\bibinfo{author}{\bibfnamefont{Y.}~\bibnamefont{Feng}},
  \bibinfo{author}{\bibfnamefont{R.}~\bibnamefont{Duan}}, \bibnamefont{and}
  \bibinfo{author}{\bibfnamefont{M.}~\bibnamefont{Ying}},
  \bibinfo{journal}{Phys. Rev. A} \textbf{\bibinfo{volume}{70}},
  \bibinfo{pages}{012308} (\bibinfo{year}{2004}).

\bibitem[{\citenamefont{Barnum et~al.}(1996)\citenamefont{Barnum, Caves, Fuchs,
  Jozsa, and Schumacher}}]{barnum96a}
\bibinfo{author}{\bibfnamefont{H.}~\bibnamefont{Barnum}},
  \bibinfo{author}{\bibfnamefont{C.}~\bibnamefont{Caves}},
  \bibinfo{author}{\bibfnamefont{C.}~\bibnamefont{Fuchs}},
  \bibinfo{author}{\bibfnamefont{R.}~\bibnamefont{Jozsa}}, \bibnamefont{and}
  \bibinfo{author}{\bibnamefont{Schumacher}}, \bibinfo{journal}{Phys. Rev.
  Lett.} \textbf{\bibinfo{volume}{76}}, \bibinfo{pages}{2818}
  (\bibinfo{year}{1996}).

\bibitem[{\citenamefont{Jozsa}(1994)}]{jozsa94a}
\bibinfo{author}{\bibfnamefont{R.}~\bibnamefont{Jozsa}}, \bibinfo{journal}{J.
  Mod. Opt.} \textbf{\bibinfo{volume}{41}}, \bibinfo{pages}{2315}
  (\bibinfo{year}{1994}).

\bibitem[{\citenamefont{Lancaster and Tismenetsky}(1985)}]{lancaster85}
\bibinfo{author}{\bibfnamefont{P.}~\bibnamefont{Lancaster}} \bibnamefont{and}
  \bibinfo{author}{\bibfnamefont{M.}~\bibnamefont{Tismenetsky}},
  \emph{\bibinfo{title}{The Theory of Matrices, 2nd edition with
  applications}}, Computer Science and Applied Mathematics
  (\bibinfo{publisher}{Academic Press, Inc.}, \bibinfo{address}{San Diego},
  \bibinfo{year}{1985}).

\bibitem[{\citenamefont{Fill and Fishkind}(1998)}]{fill98a}
\bibinfo{author}{\bibfnamefont{J.}~\bibnamefont{Fill}} \bibnamefont{and}
  \bibinfo{author}{\bibfnamefont{D.}~\bibnamefont{Fishkind}},
  \bibinfo{journal}{SIAM. J. on Matrix Analysis and Appl.}
  \textbf{\bibinfo{volume}{21(2)}}, \bibinfo{pages}{629}
  (\bibinfo{year}{1998}).

\bibitem[{\citenamefont{Anderson and Duffin}(1969)}]{anderson69a}
\bibinfo{author}{\bibfnamefont{W.~J.} \bibnamefont{Anderson}} \bibnamefont{and}
  \bibinfo{author}{\bibfnamefont{R.}~\bibnamefont{Duffin}},
  \bibinfo{journal}{J. of Math. Analysis and Appl.}
  \textbf{\bibinfo{volume}{26}}, \bibinfo{pages}{576} (\bibinfo{year}{1969}).

\bibitem[{\citenamefont{Marsaglia and Styan}(1972)}]{marsaglia72a}
\bibinfo{author}{\bibfnamefont{G.}~\bibnamefont{Marsaglia}} \bibnamefont{and}
  \bibinfo{author}{\bibfnamefont{G.}~\bibnamefont{Styan}},
  \bibinfo{journal}{Canad. Math. Bull.} \textbf{\bibinfo{volume}{15(3)}},
  \bibinfo{pages}{451} (\bibinfo{year}{1972}).

\bibitem[{\citenamefont{Du\v{s}ek et~al.}(2000)\citenamefont{Du\v{s}ek, Jahma,
  and L\"utkenhaus}}]{dusek00a}
\bibinfo{author}{\bibfnamefont{M.}~\bibnamefont{Du\v{s}ek}},
  \bibinfo{author}{\bibfnamefont{M.}~\bibnamefont{Jahma}}, \bibnamefont{and}
  \bibinfo{author}{\bibfnamefont{N.}~\bibnamefont{L\"utkenhaus}},
  \bibinfo{journal}{Phys. Rev. A} \textbf{\bibinfo{volume}{62}},
  \bibinfo{pages}{022306} (\bibinfo{year}{2000}).

\bibitem[{\citenamefont{Eldar and Forney}(2001)}]{eldar01a}
\bibinfo{author}{\bibfnamefont{Y.}~\bibnamefont{Eldar}} \bibnamefont{and}
  \bibinfo{author}{\bibfnamefont{G.}~\bibnamefont{Forney}},
  \bibinfo{journal}{IEEE Trans. Inf. Theory} \textbf{\bibinfo{volume}{47(3)}},
  \bibinfo{pages}{858} (\bibinfo{year}{2001}).

\end{thebibliography}
\end{document}